\documentclass[prb,twocolumn,showpacs,floatfix]{revtex4}
\usepackage{amsmath}
\usepackage{amssymb}
\usepackage{amsbsy}
\usepackage{graphicx}
\usepackage{verbatim}
\usepackage{bm}
\usepackage{epstopdf}

\begin{document}
\title{Ordered Loop Current States in Bilayer Graphene}
\author{Lijun Zhu,  Vivek Aji,  Chandra M. Varma}
\affiliation{Department of Physics and Astronomy, University of California, Riverside, California 92521, USA}

\begin{abstract}
While single-layer graphene shows extraordinary phenomena which are stable against electronic interactions, the non-interacting state of bilayer graphene is unstable to infinitesimal interactions leading to one of many possible exotic states. Indeed a gapped state is found in experiments but none of the states proposed so far can provide full accounts of its properties. Here we show that a {\it magnetoelectric} (ME) state  is consistent with the experimental observations.  This state breaks time-reversal symmetry through a pair of spontaneously generated current loops in each layer, and has odd-parity with respect to the two layers. We also suggest further experiments to check whether the ME state is indeed the gapped state found in experiments.  

\end{abstract}
\pacs{73.22.Pr, 72.80.Vp, 71.10.-w} 
\maketitle

\section{Introduction}
\label{sec:intro}

The non-interacting electronic state in bilayer graphene with AB (Bernal) stacking, whose lattice structure is illustrated in Fig. 1, has a pair of degenerate valence and conduction bands at  two momentum points  $K,K'=(0, \pm 4 \pi/3\sqrt{3})$ in the Brillouin zone. The energy varies quadratically with the momentum about these points, in contrast with the linear dispersion in the single-layer graphene.~\cite{McCann2006}  With hopping energy $t_{\bot}$ between the stacking carbon atoms in different layers, the other two pairs of conduction and valence bands have energy at $ \pm t_{\bot}$ at the degeneracy points. Such a state with chemical potential at the charge neutrality point is unstable to infinitesimal electron-electron interactions. In weak-coupling approximation, in which the interaction energies are small compared to $t_{\bot}$, one may look for instabilities restricting the Hamiltonian to the set of lowest energy conduction and valence bands. In such a reduced basis, a wide variety of symmetry-breaking states have been proposed as possible ground states~
\cite{Levitov2010,MacDonald2011,MacDonald2011b,MacDonald2010,Falko2010,Yang2010,Castro2008,Vafek2010,Khartonov2011,Martin2008,Vafek2011,MacDonald2011c,Vafek2012} including nematic state,~\cite{MacDonald2011,MacDonald2011b,MacDonald2010,Falko2010,Yang2010,Castro2008} anomalous quantum Hall effect (AHE) state,~\cite{Levitov2010,MacDonald2011} and layered quantum antiferromanget (AFM).~\cite{MacDonald2011,MacDonald2011b,MacDonald2010,Khartonov2011,Vafek2011,MacDonald2011c}.  However, the conductance experiments,~\cite{Lau2011,Lau2012,Yacoby2009,Yacoby2010,Freitag2011,Kim2012} done with high mobility samples and with both a top and a bottom gate to ensure that  the chemical potential is at the charge neutrality point,~\cite{Lau2011, Lau2012} show an insulating state with characteristics which are not met by the proposed states.~ \cite{footnote}

The two most important experimental findings can be summarized as follows: \\
(1) The state at charge neutrality shows a gap in two-terminal conductance measurements with a conductance $G$ smaller than the limit of measurement $\lesssim 10^{-2} e^2/h$ for voltage $V<E_{g0}$, above which 
\begin{equation}
\label{cond}
G \propto (V^2-E_{g0}^2)^{-1/2}, ~~ V \gtrsim E_{g0},
\end{equation}
with $ E_{g0} \approx \text{2 meV}$, indicating an insulating state. \\
(2) In small magnetic fields, the gap increases monotonically with a field; the conductance is consistent with a gap
\begin{equation}
\label{deltaH}
E_{g}(H) \approx \sqrt {E_{g0}^2 + \omega_c^2}, ~~ \omega_c = eB/m^*c
\end{equation}
with $m^* \approx (1/20) m$, $m$ the free-electron mass, indicating an orbital effect.

The AFM state has a gap above which the conductance has the form of Eq. (\ref{cond}), but its gap is insensitive to the external finite magnetic field in the mean-field approximation. A self-consistent calculation by interpolating the $B=0$ insulating state to a high-$B$ quantum Hall ferromagnet state could give a linear $B$ dependence of the gap at high magnetic fields~\cite{Khartonov2011,Vafek2011} (this has also been pointed out in Ref.~\cite{Levitov2010}). But the extension to low fields produces additional features which are not observed in experiments.~\cite{Vafek2011} The authors argue that this may relate to experimental limitations. The AHE state has a gap in the {\it bulk} whose variation with field is also of the form (\ref{deltaH}) (Refs.~\onlinecite{Levitov2010,Haldane1988}; see also Fig.~\ref{fig:gapvsb}). In addition, it has a surface band with a quantized Hall conductance of $4 e^2/h$. This contributes to the two probe conductance measurements and therefore is not consistent with the observations. 

\begin{figure}
\includegraphics[width=\columnwidth]{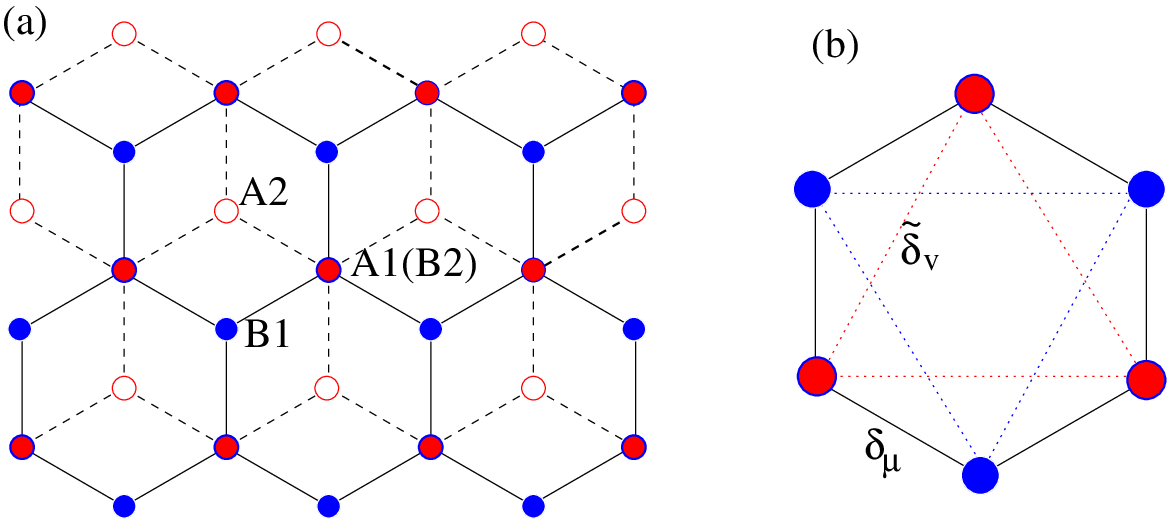}
\caption{(Color online) (a) Schematic plot of the bilayer graphene lattice (top view). Bonds in layers 1 and 2 are indicated by solid and dashed lines, respectively. Here, A atoms in layer 1 (A1) and B atoms in layer 2 (B2) are stacked on top of each other.  (b) is a unit cell in layer 1. Here $\delta_v$ and $\tilde{\delta}_v$ represent the bonds between the nearest neighbors (nn) and the next-nearest neighbors (nnn), respectively. }
\label{fig:lattice}  
\end{figure}

In this paper, we propose a magnetoelectric (ME) state, with spontaneously generated current loops from next-nearest-neighbor electron interaction ($V_{\text{nnn}}$), which has properties consistent with the above experimental observations. This state [see Figs.~\ref{fig:flux_pattern}(d)-\ref{fig:flux_pattern}(f)] breaks the time-reversal symmetry and has an odd parity with respect to two layers. As the product of time-reversal and inversion symmetries is preserved, it does not have the topologically protected surface bands as in the AHE state. In contrast, the loop current state which has an even parity with respect to two layers [see Figs.~\ref{fig:flux_pattern}(a)-\ref{fig:flux_pattern}(c)], is an AHE state.~\cite{Fradkin2008,Varma2011} However, in weak coupling, the ME state is a semi-metal with hole (electron) pockets at $K$ ($K'$) points.  We show that  when the coupling is larger ($V_{\text{nnn}} \gtrsim  t_{\bot}/5$), the ME state becomes gapped and is energetically favored over the AHE state. This regime demands the full four band basis and is beyond the weak-coupling approaches. In addition to showing that the dependence of the gap with magnetic field is consistent with the experimental observations, we also illustrate new properties of this state and propose experiments to differentiate it from other theoretical proposals.   

The rest of paper is organized as follows. We introduce a theoretical model for bilayer graphene in Sec.~\ref{sec:model}, which includes a long-range (next-nearest neighbor) Coulomb interaction term in addition to the tight-binding Hamiltonian. We show by a mean-field analysis, in Sec.~\ref{sec:meanfield}, time-reversal symmetry-breaking states with ordered loop currents arising due to the long-range interaction. They can further be classified by the inversion symmetry with respect to two layers as AHE and ME states. We further study the energetics, topological properties, as well the properties under an external magnetic field of these states by numerical analysis, in Sec.~\ref{sec:numeric}.  Some details of the calculations as well as additional results are presented in the Appendixes. 

\section{Model}
\label{sec:model}

We consider the Hamiltonian for bilayer graphene,
\begin{eqnarray}\label{ham}
H  &=& H_{1}+H_{2}+H_{12}+H_{\text{int}}, \\  \nonumber 
H_{l}& =& t\sum_{i,\delta_{\mu}}\left(a_{l i}^{\dag}b_{l i+\delta_{\mu}} \right) \\ \nonumber 
&&+ t_1\sum_{l,\tilde{\delta}_{\nu}}\left(a_{l i}^{\dag}a_{l i+\tilde{\delta}_{\nu}}+b_{l i}^{\dag}b_{l i+\tilde{\delta}_{\nu}}\right)+\text{H.c.} , \\ \nonumber
H_{12} &=& t_{\bot}\sum_{i}a_{1i}^{\dag}b_{2 i} + \text{H.c.}, \\ \nonumber
H_{\text{int}} &=& V_{\text{nnn}}\sum_{l, i, \tilde{\delta}_{\nu}} n_{l i}n_{l i+\tilde{\delta}_{\nu}},
\end{eqnarray}
where $l$ is the layer index taking the values $\left(1,2\right)$, and $i$ labels the honeycomb lattice sites. $H_l$ is the tight-binding Hamiltonian on each layer, with nearest ($t\approx3$ eV) and next-nearest ($t_1=0.1-0.3$ eV) neighbor hoppings. $(\delta_{1}$ to $\delta_{3})$ are the three vectors connecting the nearest-neighbor sites on each layer and $(\tilde{\delta}_{1}$ to  $\tilde{\delta}_{6})$ are the 6 vectors connecting the next-nearest-neighbor sites [see Fig.1(b)]. $H_{12}$ is the tight-binding part between two layers, for which we only keep the hopping between stacked atoms $t_{\bot} \approx 0.4$ eV. To focus on the time-reversal symmetry-breaking states through orbital loop currents, we only consider the next-nearest-neighbor interaction $V_{\text{nnn}}$ in $H_{\text{int}}$. $n_{l i} = a_{l i}^{\dag}a_{l i}$ or $b_{l i}^{\dag}b_{l i}$ is the charge density operator. We assume that other interactions such as onsite interaction are not strong enough to generate an order and only renormalize the parameters for the states we consider. In particular, we show in Appendix~\ref{sec:Vnn} that the nearest-neighbor interaction  
$ V_{\text{nn}}\sum_{l, i, \delta_{\mu}} n_{l i}n_{l i+\delta_{\mu}}$ need not be considered even though $V_{\text{nn}} \approx 2 V_{\text{nnn}}$. We also assume the degeneracy in spin degrees of freedom and therefore the spin index is dropped.

\section{Mean-Field Analysis}
\label{sec:meanfield}

The mean-field analysis of the model starts with the decomposition of the diagonal in the spin part of the density interactions between sites as was done to derive time-reversal breaking ME states in cuprates~\cite{CMV1999}
\begin{eqnarray}
\label{mf}
n_{ l i}n_{l, i+{\tilde \delta}_v} &=& -\left(O_{l i,i+{\tilde \delta}_v}^{\dag}O_{l i,i+{\tilde \delta}_v}/2 +n_{l i}+n_{l,i+{\tilde \delta}_v}\right), \nonumber \\
O_{l i,i+{\tilde \delta}_v} &=&   \imath a_{l i}^{\dag}a_{l i+{\tilde \delta}_v}+\text{H.c.} \text{ or }  \imath b_{l i}^{\dag}b_{l i+{\tilde \delta}_v}+\text{H.c.}.
\end{eqnarray}
The one-particle terms can be dropped. Now a mean-field approximation is made with $V_{\text{nnn}} \langle  O_{l i,i+{\tilde \delta}_v}\rangle \equiv r$. This leads to a mean-field Hamiltonian in tight-binding form but with complex hopping between next-nearest-neighbor sites:
\begin{equation}
\label{orderparam}
 ({t}_1+ir) \sum_{l,\tilde{\delta}_{\nu}}\left(a_{l i}^{\dag}a_{l i+\tilde{\delta}_{\nu}}+b_{l i}^{\dag}b_{l i+\tilde{\delta}_{\nu}}\right)+\text{H.c.} ,
\end{equation}
as well as an energy term $r^2/2V_{\text{nnn}}$. $r$ is determined by minimization of energy.

\begin{figure}[t]
\includegraphics[width=0.8\columnwidth]{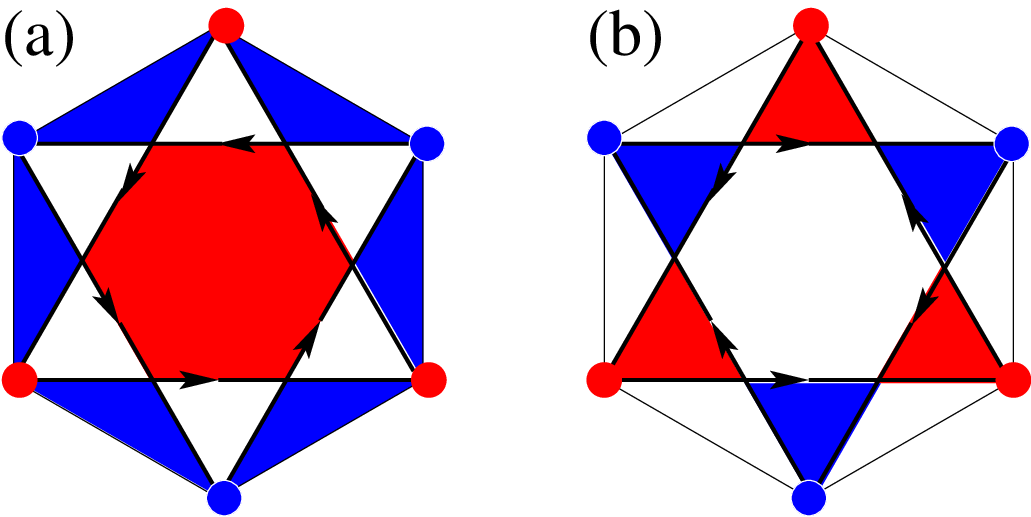}
\caption{(Color online) The pattern of net fluxes in one layer (the top layer is illustrated as an example). The red and blue colors represent the positive (counterclockwise) and negative (clockwise) fluxes, respectively. For each layer, there are two possible loop ordered states generated through $V_{\text{nnn}}$: the triangles enclosed by bonds connecting A1 and B1 sublattices and with the same center have the same (a) or opposite (b) signs of fluxes. (a) corresponds to the Haldane AHE flux pattern for a single-layer honeycomb lattice (Ref.~\onlinecite{Haldane1988}). Here, the inversion symmetry is preserved because the center of the honeycomb lattice is the inversion point. However, the inversion symmetry is always broken for (b).}
 \label{fig:flux_onelayer}
\end{figure}

\begin{figure}[tbh]
\includegraphics[width=\columnwidth]{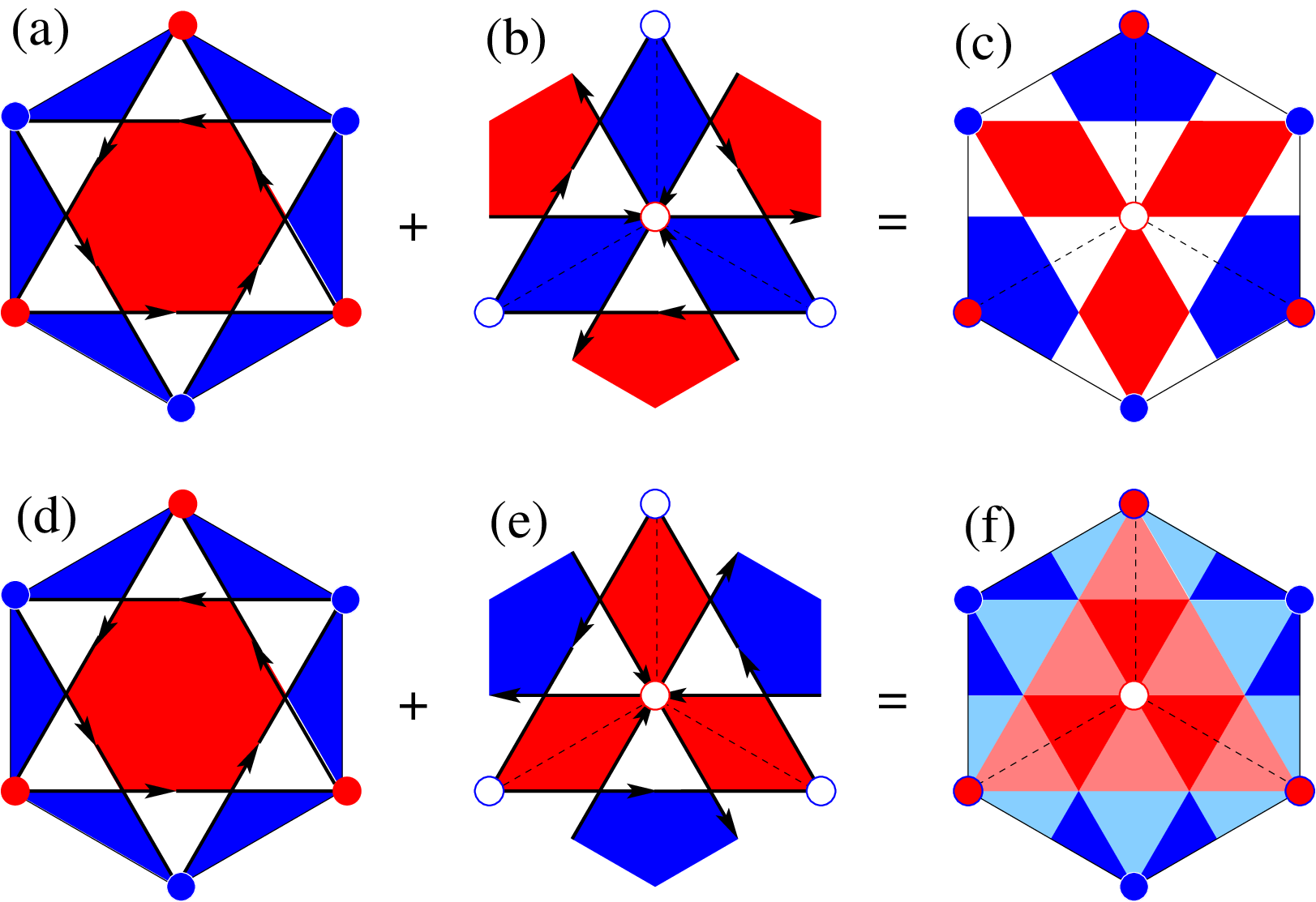}
\caption{(Color online) The pattern of fluxes in the bilayer as viewed from above. (a)-(c) represent the AHE state, while (d)-(f) represent the ME state, as discussed in the text. For each state, the flux patterns for the top layer [(a) and (d)], the bottom layer [(b) and (e)], and the net fluxes [(c) and (f)] are shown. The color representation is the same as in Fig~\ref{fig:flux_onelayer}. In (f), we use light red (blue) to represent weaker strengths of fluxes compared to the regions with regular red (blue) colors.}
 \label{fig:flux_pattern}
\end{figure}

When $r$ is finite and has the same sign (loop currents) along the triangular loop of the next-nearest-neighbor bonds for one sublattice, an ordered flux pattern is formed with alternating orientations (signs) between neighboring enclosed triangular areas (see Fig.~\ref{fig:flux_onelayer}). Classified by the sign combinations of two sublattices on two layers, four kinds of ordered loop current states can be generated through $V_{\text{nnn}}$ which break time reversal without breaking translational symmetry. (1) Within each layer, the triangles of the A sublattice and the B sublattice have the same or opposite signs of flux (see Fig.~\ref{fig:flux_onelayer}). (2) Between two layers, the triangles of the unstacked atoms (B1 and A2) centered at the stacked atoms (A1 and B2) have the same or opposite signs of flux as well [see Figs.~\ref{fig:flux_pattern}(a), \ref{fig:flux_pattern}(b), \ref{fig:flux_pattern}(d), and \ref{fig:flux_pattern}(e)]. 
Of these the two states which break inversion through having flux in the stacked and unstacked triangle of atoms in a given layer in opposite direction [Fig.~\ref{fig:flux_onelayer}(b)] always have a higher ground-state energy than the other two constructed from Fig.~\ref{fig:flux_onelayer}(a) and will not be considered further.  
We are then left with two possibilities, with flux in the stacked and unstacked triangle of atoms in a given layer in the same direction. They can be further classified by the second condition, and are \\
(i) The Haldane or AHE state of the bilayer in which the orientations of the flux in the loops atop each other in the two layers are the same. The flux patterns are shown in Figs.~\ref{fig:flux_pattern}(a)-\ref{fig:flux_pattern}(c). \\
(ii) The ME phase with the opposite orientations of the flux in the loops atop each other in the two layers, Figs.~\ref{fig:flux_pattern}(d)-\ref{fig:flux_pattern}(f). In both cases the net flux through a unit cell is zero. 

We calculate numerically the ground-state energy for the mean-field Hamiltonian in the space of all four bands. We assume the magnitude of $r$ to the same for both stacked and unstacked atoms in each layer. Taking $t_{\bot}=0.2t$ and $t_1=0$, we find the leading dependence in energy at zero temperature on the order parameter $r$ for the AHE and the ME states to be, respectively,
\begin{eqnarray}
E(r)_{\text{ME}} /t &=&  \left( \frac 1{2V_{\text{nnn}}} - 9.056 \right) r^2 + 228.5 r^4, \nonumber \\ 
E(r)_{\text{AHE}} /t &=&  \left( \frac 1{2V_{\text{nnn}}} - 9.422 \right) r^2 + 306.4 r^4 . 
\label{eq:gsenergy}
\end{eqnarray}
Equation~\ref{eq:gsenergy} gives the interesting result that for $V_{\text{nnn}} \gtrsim 0.05 t$, where the saddle-point values of $r$ are finite for both states, the ME state has a lower ground state energy in such mean-field calculation. As the critical value of $V_{\text{nnn}}$ is smaller for the AHE state, it also suggests that as temperature is decreased, the first state to arise through a transition with Ising symmetry is the AHE state (with a first order transition), and then to the ME state at a lower temperature. 
We calculate  the temperature dependence of the ground-state energy, from which we estimate the transition temperature to the ME state is $T_c\sim 10^{-2}t$ (for $V_{\text{nnn}}=0.1t$). The transition temperature to the AHE state is $5\%$ higher. The value of the transition temperature is expected to be depressed from these 
estimates when fluctuations are included.

\section{Numerical analysis}
\label{sec:numeric}

\begin{figure}[tph]
\includegraphics[width=\columnwidth]{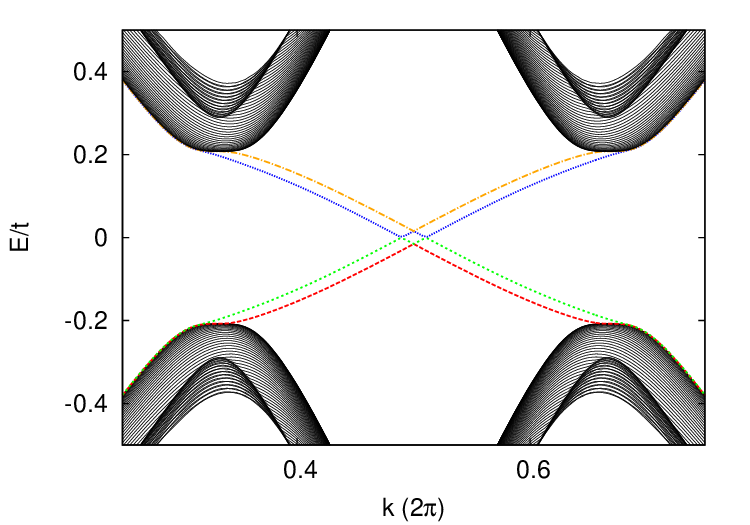}
\includegraphics[width=\columnwidth]{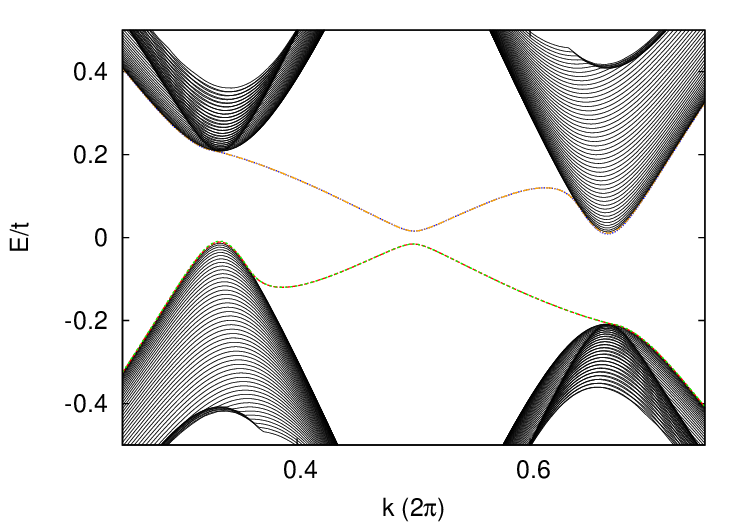}
\caption{The bandstructures with edge modes (assuming an open boundary condition along the zigzag edge) for the AHE (top panel) and ME (bottom panel) states. For the non-interacting bilayer graphene, there are  four bands: two low-energy valence and conduction bands touch each other at two momentum points $K$ and $K'$. The other two high-energy bands are split with a scale of $t_{\bot}$.  
In the AHE state, band gaps are opened at both $K$ and $K'$ due to the order parameter $r$, $\Delta \approx 6\sqrt{3} r$. In the ME state, there are also direct band gaps at $K$ and $K'$,  $\Delta_d \approx 3\sqrt{3} r$. But the center of the direct gaps shift in different directions of energy for $K$ and $K'$. When $\Delta_d \gtrsim t_{\bot}$, an indirect gap opens. For illustration purposes, we take $t_{\bot}=0.2t$ and $r=0.04t$ here, which are slightly bigger than realistic values.}
\label{fig:bandstructure}
\end{figure}

\begin{figure}[tph]
\includegraphics[width=\columnwidth]{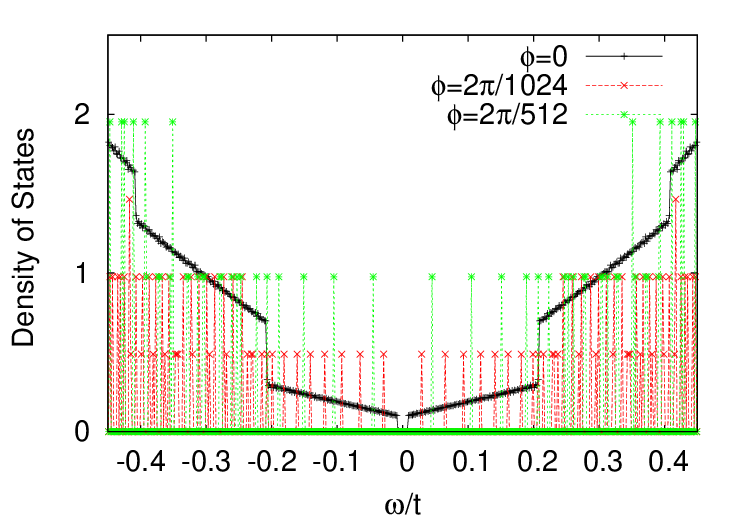}
\caption{(Color online) The density of states of the ME state without and with a magnetic field. Here $t_{\bot}=0.2t$ and $r=0.04t$. The zero-field DOS has been multiplied by a factor 5. The magnetic field strength is specified by the flux per unit cell $\phi=2\pi/M$, where $M$ is an integer, proportional to the lattice size.}
\label{fig:dos_me}
\end{figure}

We proceed to examine the properties of ME and AHE states from band structures, density of states, and topological structures. The calculational details are provided in Appendixes~\ref{sec:nummethod} and~\ref{sec:addresults}.  The bandstructures of these two states are shown in Fig. \ref{fig:bandstructure}. For the AHE state, any finite order parameter $r\propto V_{\text{nnn}}$ will open up a gap $\Delta \approx 6\sqrt{3} r$. As indicated by the crossing behaviors of the edge modes, it has a quantized Hall conductance at zero magnetic field. We verify this by calculating the Chern numbers for the four bands, which are $C=(0,2,-2,0)$ in sequence of energy for $r <O(t_{\bot})$ [for $r >O( t_{\bot})$, the Chern numbers are $C=(2,0,0,-2)$ instead]. This gives a quantized Hall conductance $\sigma_{xy} = 4 e^2/h$.  While there is a direct gap at $K$ (or $K'$) for infinitesimal $r$, the ME state has an indirect gap (see the bottom panel of Fig. \ref{fig:bandstructure}) only for $r>O( t_{\bot})$.  For small $r$, one obtains a semi-metal. The insulating ME state is not accessible within the two-band low-energy approximation.  As shown in Fig. \ref{fig:bandstructure}, the edge modes do not cross; so the ME state does not carry a net Hall current at zero field. This is again verified by the Chern number calculations.  
The density of states (DOS) of the ME state is shown in Fig.~\ref{fig:dos_me}. The vanishing DOS near $\omega=0$ is associated with the indirect bandgap.  We expect that the density of states of the ME state at the edge of the gap becomes much sharper because the excitonic effects are not included in our calculation.  
There are two jumps of DOS at high energies ($\omega\approx 0.2t$ and $0.4t$ in Fig.~\ref{fig:dos_me}).  This feature is associated with the direct band gap which is the order of $t_{\bot}$ and is a signature of the ME state. 

\begin{figure}[tph]
\includegraphics[width=\columnwidth]{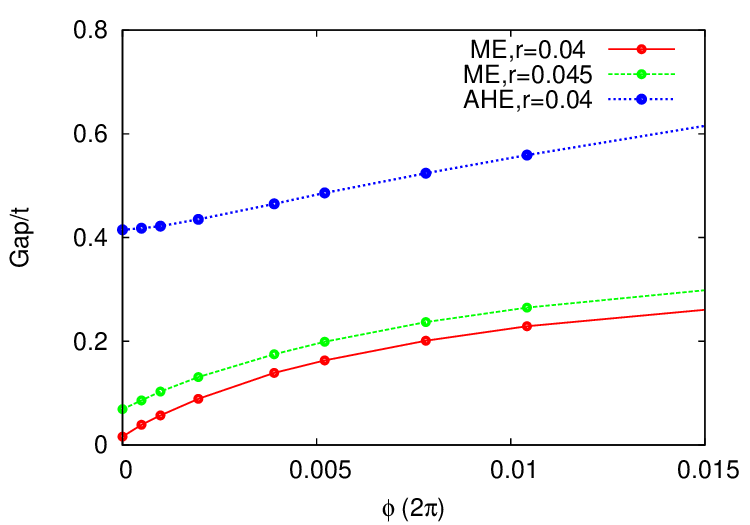}
\caption{(Color online) The gap as a function of the external magnetic field. Here $t_{\bot}=0.2t$. $B\propto \phi=2\pi/M$, and $M$ has been chosen as $128$, $256$, $\ldots$, $2048$.}
\label{fig:gapvsb}
\end{figure}

We now examine the property of the ME state in a finite magnetic field. The density of states is shown in Fig.~\ref{fig:dos_me}. Unlike the AHE state, where the lowest Landau level is pinned to the edge of the gap,~\cite{Haldane1988} the lowest Landau level(s) shifts up in energy, leading to a larger gap.  The magnetic field dependence of the gap is shown in Fig.~\ref{fig:gapvsb}, and agrees with experiments when extrapolated to low fields of the experiments.   

\section{Conclusions}
\label{sec:conclusion}

In summary, we have shown that a magneto-electric state, with loop currents ordered in each layer as in Haldane's phase but with odd-parity combination of two layers, is the ground state of interacting bilayer graphene. It has a gap which increases monotonically with the applied magnetic field, which is in agreement with the experimental observations on a gapped state in bilayer graphene. 

We have also considered the possibility of the actual realization of the Haldane AHE state in a single layer graphene by our mean-field procedure. Again the nearest-neighbor interactions are ineffective and need a very large value to get any change in symmetry, but a next-nearest-neighbor interaction $V_{\text{nnn}}$ larger than about 0.1$t$ leads to the Haldane state. This requires a dielectric constant to be smaller than about 10, which appears to be close to what experiments report.

We discuss further experiments which test the applicability of our proposed ME state, to differentiate from other proposed states from weak-coupling. (1)  The ME state is characterized by not only an indirect gap (across $K$ and $K'$ points), which is related to the observed gap $\sim 2$ meV, but also a direct gap (at $K$ or $K'$ point), proportional to the loop current order parameter, of the order of the interlayer hopping. This is different from other weak-coupling proposals where only a low-energy gap exists. The direct gap leads to sharp features at high energies, such as the  jumps in the density of states around $t_{\bot}$, as shown in Fig.~\ref{fig:dos_me}. We suggest optical absorption measurements on high mobility, dual-gated suspended bilayer graphene samples to check these features. (We note, however, previous optical absorption measurements of bilayer graphene on a substrate, which do not realize an insulating state, have not observed a direct band gap at the nodes.~\cite{optical})  (2) The ME state breaks not only time-reversal symmetry through loop current patterns in each layer, but also inversion symmetry across the layers. The former leads to a monotonically increasing gap in a magnetic field similar to the AHE state which has even parity across the layers. We note that an inversion symmetry-breaking term, such as an electric field perpendicular to the layers, introduces a linear coupling of AHE and ME order parameters ($\sim E_{\bot} \psi_{\text{ME}}\psi_{\text{AHE}}$). This promotes the AHE state when the ME order parameter is finite. The two-terminal probe indeed showed a finite conductance $\sim 4e^2/h$ when a finite perpendicular electric field is applied,~\cite{Lau2011,Lau2012} which may be interpreted as a signature of the AHE state.
We suggest a four-terminal experiment to verify this mechanism, and therefore, provide further proof for the existence of the ME state at charge neutrality.

\begin{acknowledgements}
We acknowledge extensive discussions with C. N. Lau. L.Z. also acknowledges discussions with D.-N. Sheng. This research was partially supported by NSF Grant No. DMR-0906530 (L.Z. and C.M.V.), and UCR initial complement(V.A.).   
\end{acknowledgements}

\appendix

\section{Ordered loop current states generated through $V_{\text{nn}}$}
\label{sec:Vnn}

\begin{figure}[htp]
\includegraphics[width=0.8\columnwidth]{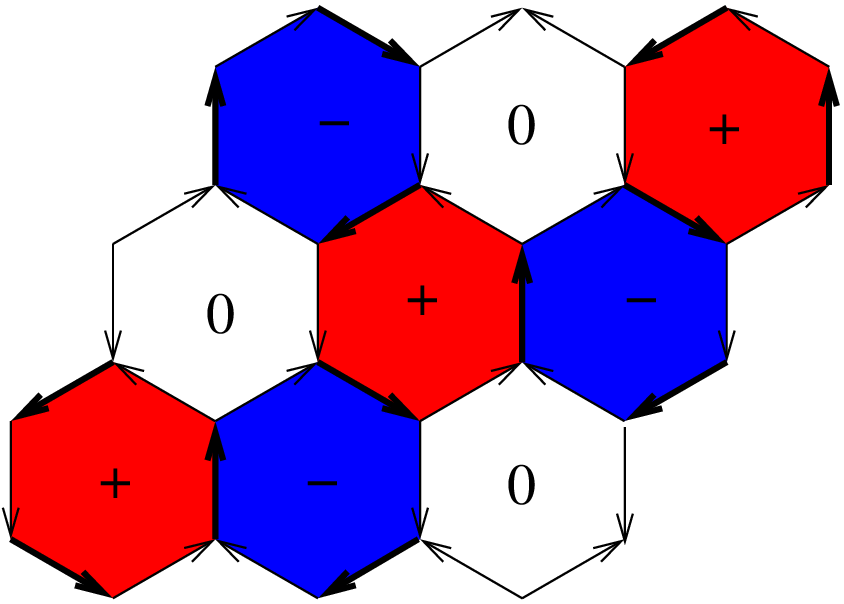}
\caption{A possible ordered loop-current state generated through $V_{\text{nn}}$. Only one layer is shown.  The red, blue and white colors represent the net positive (clockwise), negative (anticlockwise), and zero fluxes in a hexagonal cell.}
 \label{fig:kekule}
\end{figure}

For the nearest-neighbor interaction $V_{\text{nn}}$, the generated loop currents flow around the perimeter of a hexagonal cell with net flux.  If we allow flux (positive or negative) in any unit cell, we run into the problem of frustration of an Ising model in a triangular lattice. This in general is not a possible ground state because unlike with spins, alternate arrangements with lower energy are possible. The simplest is a $\sqrt{3}a \times \sqrt{3}a$ structure of hexagons depicted in Fig.~\ref{fig:kekule} in which a hexagonal cell with zero net flux is surrounded by six hexagonal cells, each with a Kekule pattern of currents in the links with alternate net positive and negative flux.  This state breaks the translational symmetry while the unit cell is enlarged to include $3\times3$ unit cells of a translational invariant honeycomb lattice.

We assume that the flux pattern in the top layer follows Fig.~\ref{fig:kekule}, which takes a sequence ($+$,$-$,$0$) from left to right.  The bottom layer could follow the same pattern, or takes another pattern ($-$,$+$,$0$). Therefore, there are two types of ordered loop current states (the other combinations are equivalent to either of these two states by rotation). We carry out a mean-field calculation on the ground-state energy for these states, and find that the states with different patterns for two layers are lower in energy.  This lower energy state is allowed only if $V_{\text{nn}} \gtrsim 2.06 t$, i.e., be comparable to the bandwidth. The reason is that such a state does not have a ``nesting" periodicity and therefore does not use any singularity in the joint density of states. 

\section{Numerical Methods}
\label{sec:nummethod}

The mean-field Hamiltonian of the ordered loop current states can be readily diagonalized for each momentum point ($k_x$,$k_y$). However, to examine their properties under an applied magnetic field, as well as the topological properties such as the edge modes, we also carry out a real-space calculation. 

\begin{figure}[tph]
\includegraphics[width=0.8\columnwidth]{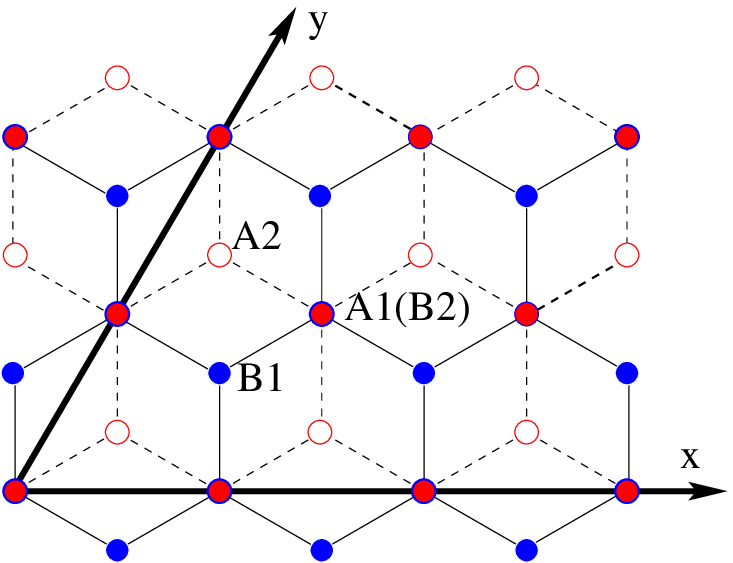}
\caption{Real-space representation of the bilayer graphene lattice. With the $x$-$y$ axis specified in the figure, the lattice can be represented by a $N_x \times N_y$ rectangle lattice with four atoms in each lattice site.}
\label{fig:lattice}
\end{figure}

Following a common numerical practice for honeycomb lattices, we describe the bilayer graphene by a $N_x \times N_y$ lattice, with four atoms on each lattice site (see Fig.~\ref{fig:lattice}). The real-space coordinations of each atom (and the momentum) in the orthogonal $x$-$y$ axis can be easily obtained from this ``deformed''-lattice representation.  

The effect of an external magnetic field can be captured by imposing a phase to the hopping term to each bond $t_{ij} \to t_{ij} e^{i\int {\bf A} \cdot d{\bf x}}$.  We adopt the Landau gauge $A_y = - B x$ to take advantage of the translational invariance along the $y$ direction.  For the periodic boundary condition, we commonly take the flux due to the external magnetic field in a unit cell to be $\phi = 2\pi /M$, where $M$ is an integer proportional to $N_x$. $M=1000$ is equivalent to a magnetic field strength $B \approx 30T$ for graphene systems. 

In general, we choose periodic boundary conditions along both directions ( a torus). This provides a verification of our momentum-space calculations for the zero field.  For the edge modes calculation, we choose an open boundary along the $x$ axis, which corresponds to a cylinder with the zigzag edge.

\section{Additional results}
\label{sec:addresults}

\subsection{Density of states}

\begin{figure}[tph]
\includegraphics[width=\columnwidth]{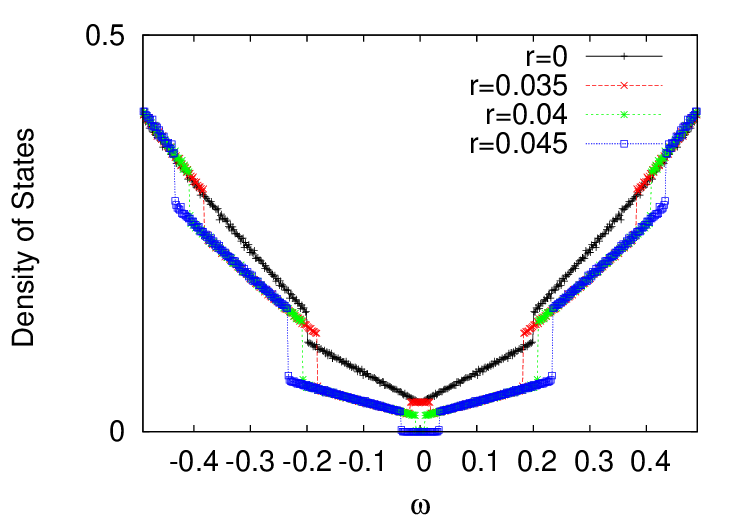}
\caption{The density of states for the ME state. Various values of the loop current order parameter are shown. The non-interacting case $r=0$ is also shown in comparison. Here $t_{\bot}=0.2t$.}
\label{fig:dos}
\end{figure}

In Fig.~\ref{fig:dos}, we show the DOS for the ME state with various values of the order parameter $r$. We also show the result for the non-interacting case as well. For $r=0$, the DOS at $\omega=0$ is finite, which is due to the parabolic dispersion $E\sim k^2$.  There is a sudden jump of DOS at $\omega\approx \pm t_{\bot}/2$, which is the energy scale of the gap between the top two valence bands (bottom conduction bands). When $r$ is small, the conduction band at the $K$ point and the valence band at the $K'$ point have overlaps in energy. Therefore, it remains a semimetal state. A full gapped state happens when $ r \gtrsim t_{\bot}/(3\sqrt{3})$. 

\subsection{Chern numbers}

In the main paper, we have shown the topological properties of the AHE and ME states by showing their edge modes, calculated by assuming a periodic boundary condition along the $y$ axis and an open boundary condition along the $x$ axis.  Judging from the crossed edge modes, we learn the AHE state indeed has a finite Hall conductance at zero field. Similarly, the ME state does not have any Hall effect at zero field.

This topological property can also be checked from the Chern numbers. The Chern number for a given band is defined as 
\begin{equation}
C_n=\int {d^2 k \over (2\pi)^2} {\hat z} \cdot {\partial \Psi_n^*({\bf k}) \over \partial k_x} \times  {\partial \Psi_n({\bf k}) \over \partial k_y},
\label{eq:chern}
\end{equation}
 where $\Psi_n({\bf k})$ is the wavefunction for the $n$-th band. The Hall conductance is given by $\sigma_{xy} = Ce^2/h$. 
We calculate the Chern numbers in a momentum-space algorithm, by summing the Berry curvature in small areas $(\Delta k_x, \Delta k_y)$.

For the AHE state, any finite order parameter $r$ will open up a gap. We find the Chern numbers for the four bands are $C=(0,2,-2,0)$ in sequence of energy, i.e, the valence and conduction bands away from the Fermi energy do not carry any Chern numbers.  However, when $r \gtrsim t_{\bot}$, the Chern numbers become $C=(2,0,0,-2)$ instead. 
In either case, a finite Hall conductance is a robust feature. 

For the ME state,  an indirect gap is open only for a finite order parameter $r\gtrsim t_{\bot}$. However, as long as the four bands do not touch each other (having direct band gaps), the Chern number calculation is robust. We find the Chern numbers for the four bands of the ME state are all zero, indicating a non-quantum Hall state. When $r \gg t_{\bot}$, we find the Chern numbers become $C=(1,-1,1,-1)$. In this limit, each layer is a Haldane state while $t_{\bot}$ only acts as a perturbation, slightly splitting the states from two layers. Still, there is no Hall effect near the Fermi energy. 

\subsection{Edge modes in a finite magnetic field}

\begin{figure}[tph]
\includegraphics[width=\columnwidth]{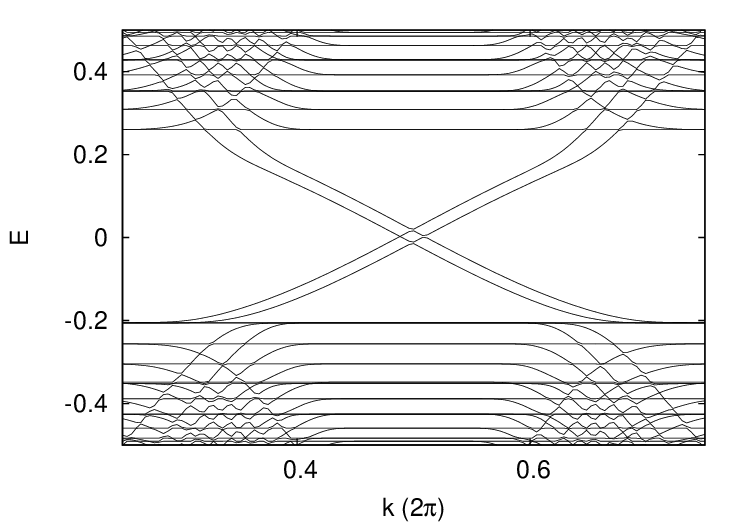}
\includegraphics[width=\columnwidth]{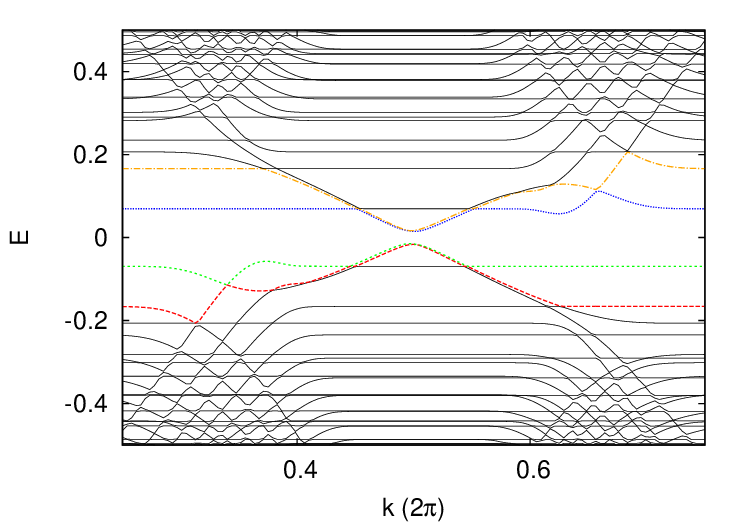}
\caption{The edge modes for AHE (top panel) and ME (bottom panel) states at a finite magnetic field. Here $t_{\bot}=0.2t$ , $r=0.04t$, and $\phi=2\pi/256$.}
\label{fig:edgemodeB}
\end{figure}

In the main text, we have shown the edge modes for the AHE and ME states at zero magnetic field. For completeness, we also show the edge modes in a finite external magnetic field, in Fig.~\ref{fig:edgemodeB}. 

\end{document}